\documentclass[aps,pre,preprint,showpacs]{revtex4-1}
\usepackage{graphicx}
\usepackage{amsmath}
\usepackage{amsfonts}
\usepackage{amssymb}
\usepackage{amsthm}

\begin{document}

\title{Optimal random search for a single hidden target}

\author{Joseph Snider}
\email{j1snider@ucsd.edu}
\affiliation{The University of California, San Diego, CA}
\date{\today}

\begin{abstract}
A single target is hidden at a location chosen from a predetermined probability distribution. Then, a searcher must find a second probability distribution from which random search points are sampled such that the target is found in the minimum number of trials. Here it will be shown that if the searcher must get very close to the target to find it, then the best search distribution is proportional to the square root of the target distribution regardless of dimension. For a Gaussian target distribution, the optimum search distribution is approximately a Gaussian with a standard deviation that varies inversely with how close the searcher must be to the target to find it. For a network, where the searcher randomly samples nodes and looks for the fixed target along edges, the optimum is to either sample a node with probability proportional to the square root of the out degree plus one or not at all.
\end{abstract}

\pacs{02.50.-r 02.60.-x 89.75.Hc}

\maketitle

\section{Introduction}

Imagine a single target hidden with some known distribution and a searcher trying to locate it by randomly drawing points from another distribution. Assuming the cost for the search only depends on the number of steps, an optimal searcher has to choose a distribution that minimizes the average number of steps to locate the target. This situation is different from the usual search paradigms in which there are many targets\cite{viswanathan1999,marell2002,sims2008,humphries2010} and/or no knowledge of where the target may be\cite{benichou2005,reynolds2008}. When there are many targets, the solution is to perform a random walk with a L\'{e}vy flight, but that depends on expanding the search time in the distance between targets which is not possible when there is only one target\cite{viswanathan1999}. Intermittent searches are optimal if there is no knowledge of the target and large steps cost more than small ones\cite{pmid16090215,benichou2005,slutsky2004,eliazar2007}.

The goal here is to describe the optimum guess distribution to sample to find a target, given the target distribution and assuming the searcher must be within a distance $R$ of the target to consider it found and with no additional cost for large steps. A familiar example of this is searching with the eye for a particular target\cite{najemnik2008,najemnik2005}, such as a face in a crowd. Without any guess this is a very difficult proposition, but if someone who knows where the target face is gestures, even inaccurately, toward the target, then that clue greatly improves the search. In the case of the eye, the movement is very rapid compared to identification of the target\cite{oteromilan2008} so that the search time is dominated by the number of places checked rather than, for example, the total path length traversed. Also, the searcher may not see the target on the first pass, so memory is not really important. In other words, actual identification of the target is unreliable so there is no reason to avoid a spot just because the search failed there a few times. Under these assumptions, it will be shown that the optimal search distribution is different than the target distribution, and specifically, if $R$ is small compared to the variability of the target distribution, then the search distribution is proportional to the square root of the target distribution.

\section{Continuous Systems}

Now consider definitions of the relevant quantities for a continuous target distribution. Let $G(x)$ be the unknown guess distribution, $T(x)$ the known target distribution, and $R$ the range over which the guess is considered to find the target. For example, if $T(x)$ is a delta function so that the target is at $x=0$, then the best guess is clearly also $G(x)=\delta(x)$ when $R=0$. For non-zero $R$ a guess anywhere on the interval $[-R,R]$ is considered a hit, so the best guess distribution may be a boxcar from $[-R,R]$. For simplicity concentrate on one dimension, although generalizing to two (or more) is trivial.

For a fixed target at position $x$ and assuming the guesses are uncorrelated (memoryless), the average number of guesses required before finding the target is just the inverse of the probability that a guess is correct. Averaging over all possible target locations gives
\begin{equation}
   \left<n\right> = \int_{-\infty}^{\infty}dx\,\frac{T(x)}{\int_{x-R}^{x+R}dt\,G(t)}.
   \label{eq:mean_n}
\end{equation}
Note that the the denominator in the integral over $T(x)$ is non-local which is the source of some mathematical difficulties.

The optimum $\left<n\right>$ is found by taking the functional derivative of $G$ subject to the constraint that $G$ be normalized:
\begin{equation}
   0 = \frac{\delta}{\delta G(y)}\left(\left<n\right>\left[G(t)\right] 
         - \alpha\left(1-\int_{-\infty}^{\infty}dt\;G(t)\right)\right),
\label{eq:optimization_eq}
\end{equation}
where $\alpha$ is a Lagrange multiplier. To handle the non-local integral, define a function $Q(t;x,R)$ such that
\begin{equation}
   \int_{x-R}^{x+R}dt\,G(t) = \int_{-\infty}^{\infty}dt\,G(t)Q(t;x,R),
\label{eq:boxcar}
\end{equation}
\textit{i.e.}, $Q(t;x,R)$ is a boxcar function of width $2R$ centered at $x$. Then, by the generalized chain rule for functional derivatives, or as may be verified from the definition of a functional derivative
\begin{equation}
\begin{split}
   \frac{\delta}{\delta G(y)}\left<n\right> 
     &= \lim_{\epsilon\rightarrow 0}\frac{1}{\epsilon}\left(
         \left<n\right>\left[G(t)+\epsilon\delta\left(t-y\right)\right]
              - \left<n\right>\left[G(t)\right]\right)\\
      &= -\int_{-\infty}^{\infty}dx\,\frac{T(x)Q(y;x,R)}
              {\left[\int_{-\infty}^{\infty}dt\,G(t)Q(t;x,R)\right]^{2}} .\\
\end{split}
\end{equation}
By definition, $Q(y;x,R) = 1$ whenever $\left|x-y\right|<R$. Thus, $Q(y;x,R)=Q(x;y,R)$, and note that this is regardless of dimension as long as the norm is well defined. The functional derivative simplifies to 
\begin{equation}
   \frac{\delta}{\delta G(y)}\left<n\right> 
      = -\int_{y-R}^{y+R}dx\,\frac{T(x)}
              {\left[\int_{x-R}^{x+R}dt\,G(t)\right]^{2}}.
\label{eq:functional_der}
\end{equation}
Plugging equation \ref{eq:functional_der} into equation \ref{eq:optimization_eq} gives
\begin{equation}
  \int_{y-R}^{y+R}dx\,\frac{T(x)}{\left[\int_{x-R}^{x+R}dt\,G(t)\right]^{2}} = \alpha.
\end{equation}
But, $\alpha$ is a constant and this must hold for any $y$, so the integrand must be a constant, and, indeed, it must be $\alpha/2R$, but we can rely on some constant $A$ to handle normalization. Thus, solving the integrand for $G(x)$ and noting that everything is positive ($G$ and $T$ are both probability distributions),
\begin{equation}
  \int_{x-R}^{x+R}dt\,G(t) = A\sqrt{T(x)}.
\label{eq:general_optimum}
\end{equation}
Equation \ref{eq:general_optimum} provides the most general version of the main result: how the guess distribution $G(x)$ depends on the target distribution $T(x)$.

Equation \ref{eq:general_optimum} has an interesting expansion near $R\rightarrow 0$ where the integral of $G$ is approximately constant giving
\begin{equation}
   G(x) \approx \tilde{A}\sqrt{T(x)},
   \label{eq:optimum_R_small}
\end{equation}
where $\tilde{A}$ takes care of normalization. Thus, in the limit that the range becomes small the optimal guess distribution is constructed by taking the square root of the target distribution and normalizing. This is somewhat surprising because at first glance the best choice for the guess distribution would have been to match the target distribution. The square root effectively widens the search distribution and lowers the chance of sampling at the most likely position. This is advantageous because the search is completely memoryless, so the most likely target location would be oversampled in a probability matching approach, and the searcher would waste time repeatedly sampling the same site. A simple example is to imagine just two discrete target sites, one with a 99\% probability of having the target and one with 1\%. If the searcher probability matches, then the mean time to find the target is $\left<n\right> = 0.99/0.99 + 0.1/0.1 = 2$, and the searcher finds the target in one guess 99\% of the time and looks in the wrong spot an average of 100 times on 1\% of the trials. This can be improved significantly by looking at the more likely target slightly less often, say 98\% of the trials, making the mean $\left<n\right> = 0.99/0.98 + 0.1/0.2 \approx 1.5$. The improvement comes from searching at the unlikely hiding spot slightly more often, but not often enough to impact the more likely case significantly. The square root rule finds the best trade off and generalizes to more complex, continuous situations.

A nice feature to note here is that since $G$ is actually calculated, we can at least approximate the actual value for the mean number of trials $\left<n\right>$ before finding the target (equation \ref{eq:mean_n}). Thus, if the number of steps is too large, say greater than $2\left<n\right>$ as human searchers seem to use \cite{Wolfe1998}, either the assumed target distribution is likely invalid or the target is not present, and failure is recognized. Note that this is not the case for any L\'{e}vy or intermittent search for a single target because their mean search times scale with the size of the search region\cite{redner2001}, so failure is difficult to recognize for those cases.

\begin{figure}[h]
\includegraphics[width=5in]{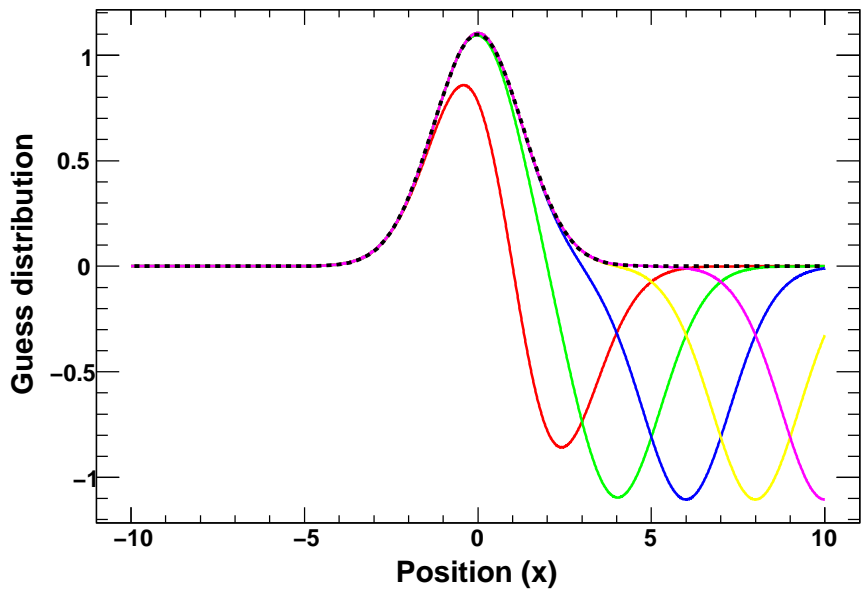}
\caption{(color online) The sum in equation \ref{eq:sum_gaussian} for a Gaussian target distribution with $\sigma=1$ and $R=1$ with the infinite sum cutoff at 1, 2, 3, 4, and 5 terms. As the number of terms is increased the negative bubble moves off to infinity leaving only the Gaussian shaped peak near the origin. The dashed black line is the best fit Gaussian ($\sigma=1.3(1)$) with 500 terms. Note that the fit Gaussian has so much error because the guess distribution is not exactly a Gaussian when $R\ne 0$. }
\label{fig:figure1}
\end{figure}

To solve for $G$ in the case when $T(x)$ is differentiable, take a derivative:
\begin{equation}
  G(x+R)-G(x-R) = \frac{A}{2}\frac{T'(x)}{\sqrt{T(x)}}.
\end{equation}
Now, redefine $x\rightarrow x-R$, $A$ to handle all the constants, and simplify the recursive equation to give
\begin{equation}
 G(x) = A\sum_{n\in\text{Odds}}\frac{T'(x-nR)}{\sqrt{T(x-nR)}},
\label{eq:g_solution_sum_form}
\end{equation}
assuming $G\left(\infty\right)\rightarrow 0$ and the sum converges.
Formally, this expression correctly gives $G$ in terms of $T$; however, there are some significant practical considerations. To understand the source of the problem, consider a Gaussian target distribution:
\begin{equation}
  T(x)\propto e^{-\frac{x^{2}}{2\sigma^{2}}}
\end{equation}
Then, plugging into equation \ref{eq:g_solution_sum_form}, $G$ is
\begin{equation}
   G(x)=A\sum_{n\in\text{Odds}}(nR-x)e^{-\frac{\left(x-nR\right)^{2}}{4\sigma^{2}}}.
   \label{eq:sum_gaussian}
\end{equation}
The problem with this expression can be seen by imagining trying to normalize $G$. Each term of the sum integrates to zero, but the whole thing must not integrate to zero since $G$ is a probability distribution. This is a problem with commuting the two infinities, in the sum and the integral, shown by numeric summation of the first few terms (figure \ref{fig:figure1}). There is a negative bubble that moves off to infinity as the number of terms increases, and a Gaussian like peak that remains near zero. Thus, the limit of the sum must be taken before that of the integral. 

Alternatively, since the negative bubble always moves to plus infinity, we can rely on the fact that $G(x)$ is even because $T(x)$ is even, and only look at $x<0$. While numerically such a process works fine, it complicates the integrals considerably as far as trying to approximate $G$ analytically.

To make some headway analytically, consider a Laplace transform approach. For convenience, define $q(x)=A\sqrt{T(x)}$, and then the bilateral Laplace transform of equation \ref{eq:general_optimum} is
\begin{equation}
\begin{split}
   \int_{-\infty}^{\infty}dxe^{-sx}\Bigg\{\int_{x-R}^{x+R}dt\,G(t) &= q(x)\Bigg\}\\
   \frac{\sinh\left(Rs\right)}{s}\tilde{G}(s) &\propto \tilde{q}(s),
\end{split}
\label{eq:laplace_transform}
\end{equation}
where the $\sinh\left(Rs\right)/s$ prefactor can be derived easily using the same boxcar trick as before (equation \ref{eq:boxcar}).

\subsection{Box car target distribution}

The Laplace transform can be very difficult to invert, so equation \ref{eq:laplace_transform} is not always useful; however, it does work for special cases of boxcar target distributions and Gaussian distributions with small $R$. First, consider the boxcar
\begin{equation}
   T(x) \propto \begin{cases}
           1&   -Q<x<Q,\\
           0&   \text{otherwise}.\\
          \end{cases}
\end{equation}

Up to a constant factor, the square root of a boxcar is also a boxcar, so
\begin{equation}
 \begin{split}
  \tilde{q}(s) &\propto \int_{-Q}^{Q}dx\,e^{-sx}\\
               &= \frac{\sinh\left(Qs\right)}{s}.
 \end{split}
\end{equation}
Then, from equation \ref{eq:laplace_transform}
\begin{equation}
   \tilde{G}(s) \propto \frac{\sinh\left(Qs\right)}{\sinh\left(Rs\right)}.
\end{equation}
For the special case $R=Q$, everything cancels and $\tilde{G}(s)\propto1$ immediately gives $G(t)=\delta(0)$ as expected. By looking at the center of the distribution the search radius covers all possible targets and is clearly optimal.

Similarly, for $R=Q/2$, we can apply the hyperbolic identity $\sinh 2x = 2\sinh x\cosh x$ to give after minor simplification
\begin{equation}
  \tilde{G}(s) \propto e^{Rs} + e^{-Rs},
\end{equation}
and the inverse is $G(x) = 1/2\left[\delta(x+R) + \delta(x-R)\right]$. Again, this perfectly tiles the search space with no overlap and is the best search strategy. All such integer fractions of $Q$ can be solved, and they similarly generate a perfectly tiled cover of the interval $[-Q,Q]$ (figure \ref{fig:figure3}B).

\subsection{Gaussian target distribution}

\begin{figure}[h]
\includegraphics[width=5in]{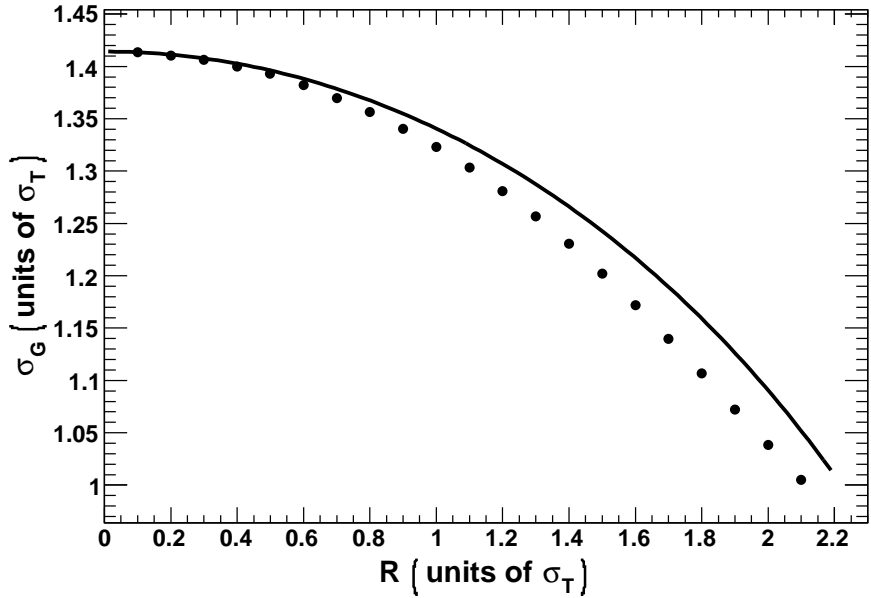}
\caption{This is the standard deviation of the optimum guess $\sigma_{G}$ versus the search radius $R$. The dots are numerically calculated by assuming the guess distribution is a Gaussian and directly optimizing $\left<n\right>$. The line is the analytical result using the Gaussian approximation for $x/\sinh x$. In both cases the target distribution is a Gaussian with units set by choosing $\sigma_T=1$.}
\label{fig:figure2}
\end{figure}

Another case of great interest is the Gaussian target distribution. In that case, notice that $x / \sinh x$ itself looks like a Gaussian, so we can approximate it as a Gaussian by matching variances. The integral of $x/\sinh x$ is standard, and the normalized second moment is $\frac{\pi^{2}}{2R^{2}}$. Thus, the the useful approximation is a Gaussian with a matching second moment
\begin{equation}
   \frac{x}{\sinh\left(Rx\right)} \approx \frac{1}{R}e^{-\frac{R^{2}x^{2}}{\pi^{2}}}.
\end{equation}
Finally, plugging in the square root of a Gaussian with standard deviation $\sigma_{T}$ for $q$ leaves (dropping all coefficients as usual)
\begin{equation}
   G(x) \propto e^{-\frac{x^{2}}{4\left(\sigma_{T}^{2}-\frac{R^2}{\pi^{2}}\right)}}
   \label{eq:gaussian_final_result}
\end{equation}
for $\sigma_{T}>R/\pi$ (figure \ref{fig:figure3}A,D). For $R/\pi>\sigma_{T}$ the approximation breaks down. Thus, the optimal guess distribution is approximately a Gaussian with standard deviation $\sigma_{G}\approx\sqrt{2}\sqrt{\sigma_{T}^{2}-\frac{R^{2}}{\pi^{2}}}$ in good agreement with numerical estimates (figure \ref{fig:figure2}).

\subsection{Convolution}

\begin{figure}[h]
\includegraphics[width=6in]{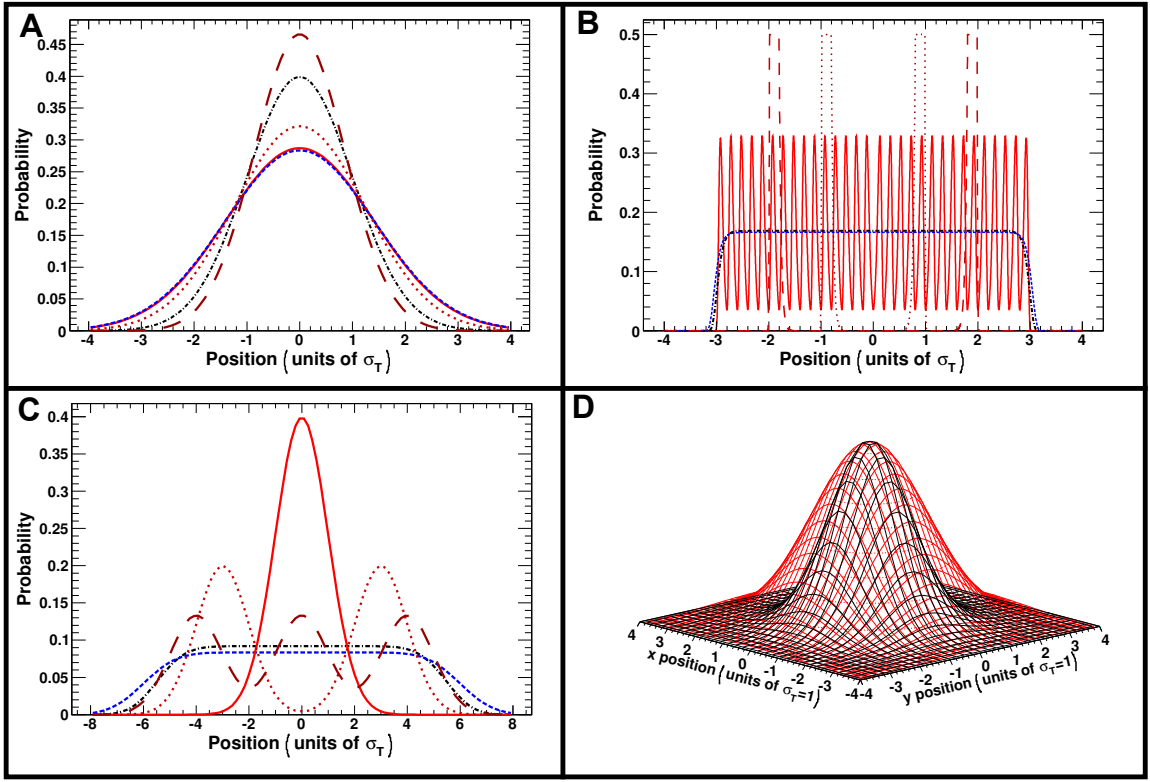}
\caption{(color online)Some interesting sample target distributions (black, dot-dashed) are shown with their associated guess distributions (red, solid, dotted, and long dashed) and the square root approximation (blue, short dashed). Part A is shows a Gaussian distribution with $\sigma_{T}=1$ to set the units and $R\in\{0.5,1.5,2.5\}$, and is calculated using equation \ref{eq:gaussian_final_result}. Part B shows a super exponential $\exp-\left(x/3\right)^{40}$ to approximate a boxcar with $R\in\{0.1,1.1,2.1\}$ calculated from the numeric sum (equation \ref{eq:g_solution_sum_form}). The optimum guess distributions approaches the delta function solutions for a true boxcar. Part C uses a target distribution of a Gaussian with $\sigma_{T}=1$ convolved with a boxcar $B_{6}(0)$. As $R\in\{2,3,6\}$ increases peaks appear, highlighting the non-local nature of the search. For $R=3$ the most likely search points are at $\pm3$, but the target is most likely at zero. Part D shows a two dimensional Gaussian target distribution (black) and the optimum guess distribution for $R=0.5$. For clarity the normalization is such that the peak heights match.}
\label{fig:figure3}
\end{figure}

As a further example highlighting the non-local nature of the search, consider the special case where the target distribution can be written in the form of a convolution squared:
\begin{equation}
 T(x) \propto \left[\int_{-\infty}^{\infty}d\tau g(\tau)f(t-\tau)\right]^{2},
\end{equation}
where $f$ and $g$ are arbitrary functions such that $T$ is a probability distribution. Then, since the Laplace transform of the convolution is the product of the Laplace transforms,
\begin{equation}
  G(s) \propto \frac{s}{\sinh\left(Rs\right)}\tilde{g}(s)\tilde{f}(s).
\end{equation}
Now we can combine the boxcar and Gaussian functions from before to highlight the interesting, non-local behavior. Let $g$ be a boxcar of width $nR$ for $n>0$ an integer as before. Then,
\begin{equation}
\begin{split}
  G(s) &\propto \frac{\sinh\left(nRs\right)}{\left(Rs\right)}\tilde{f}(s)\\
       &= \sum_{i}e^{Rx_{i}}\tilde{f}(s),
\end{split}
\end{equation}
where the sum is over $x_{i}=R+2Ri-nR$ and $i\in\mathbb{Z}_{n}$ to tile the interval from $[-nR,nR]$ evenly with subintervals of length $2R$. Next, $\exp(as)$ is the Laplace transform of $\delta(x-a)$ so
\begin{equation}
   G(x) \propto \sum_{i}f\left(x-x_{i}\right).
\end{equation}
This can lead to some interesting effects. For example, set $f\propto\exp{-\frac{x^{2}}{2\sigma^{2}}}$ and let $B_{Q}(x)$ be a boxcar of radius $Q$ centered at $x$, then the target function is
\begin{equation}
 \begin{split}
  T(x) &\propto \left[\int_{-\infty}^{\infty}d\tau B_{Q}(t-\tau)
                              e^{-\frac{\tau^{2}}{2\sigma^{2}}}\right]^{2}\\
       &= \left[\int_{t-Q}^{t+Q}d\tau e^{-\frac{\tau^{2}}{2\sigma^{2}}}\right]^{2}\\
       &\propto \left[\text{erf}\left(\frac{Q-x}{\sqrt{2}\sigma}\right) +
                      \text{erf}\left(\frac{Q+x}{\sqrt{2}\sigma}\right)\right]^{2}
 \end{split}
\end{equation}
which looks like to a Gaussian with a peak at zero. The guess distribution, on the other hand, has two peaks at $\pm R$, and the most likely place to search does not coincide with the most likely target location (figure \ref{fig:figure3}C).

\section{Discrete systems}

While the search technique introduced here is very powerful for continuous systems, it also applies to discrete systems for which the target can only be found by looking at its site. These can be considered a special case of continuous systems, or more simply, write the discrete version of equation \ref{eq:mean_n} for a set of discrete sites $\mathbb{A}$:
\begin{equation}
 \left<n\right> = \sum_{i\in\mathbb{A}}\frac{t_{i}}{g_{i}},
\end{equation}
where $t_{i}$ are the target probabilities and $g_{i}$ the guess probabilities. Then, as one could guess from the general result for small search radius (equation \ref{eq:optimum_R_small}), the optimum choice of $g_{i}$ given $t_{i}$ is $g_{i}=A\sqrt{t_{i}}$ where $A=1/\sum_{i}\sqrt{t_{i}}$ maintains normalization.

\begin{figure}[h]
\includegraphics[width=4in]{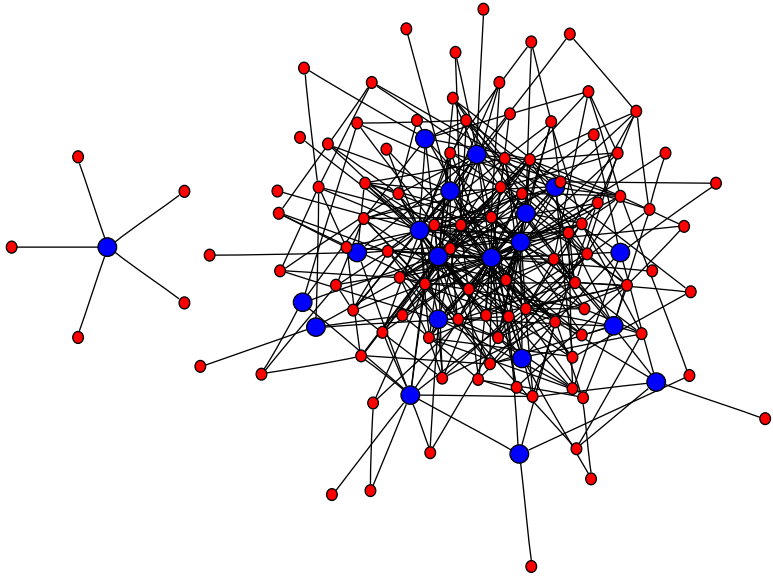}
\caption{(color online) These are two sample graphs. On the left, is a simple 5-star graph, and the obvious best strategy is to look at the center node (large, blue) where any target is found in exactly one step. On the right is the adjnoun network\cite{newman-network2006} with nodes to be looked at labeled blue, larger, and ones that should be ignored in red, smaller. The blue nodes are found using the heuristic algorithm described in the main part.}
\label{fig:figure4}
\end{figure}

A discrete set of isolated sites is quite simple, but a system of great interest and richness is networks\cite{albert2002,newman2008}. On networks, one can imagine a searcher with random access to the nodes that is able to look at any node and out along the node's edges to its nearest neighbors. This is akin to typing in a web address and checking the links on the resulting page for the target, rather than just following links from page to page. In that case, the probability of finding a single target at any given node $u$ depends on the in-degree of the node, for each of its neighbors, plus one, for the node itself: $d_{u}+1$, and the mean search time for a target randomly hidden on some fixed node is 
\begin{equation}
\left<n\right> = \sum_{i}\frac{1}{g_{i}+\sum_{j \rightarrow i}g_{j}},
\label{eq:graph_mean_search}
\end{equation}
where $i$ is a node, and the sum in the denominator is over all nodes that have an edge pointing to $i$. The mean field assumption here is that the probability that the target is hidden at node $i$ is equal to the probability of finding the target at node $i$ so that $g_{i}\propto\sqrt{1+d_{\text{out},i}}$, but it can fail spectacularly for graphs. As a simple example, consider a star graph with one node connected to 5 independent nodes (figure \ref{fig:figure4}, left), and hide the target on that graph with equal probability at each node. Then, the probability of finding the target by looking at the center node is 1, but the probability the target is hidden at the center node is only 1/6.

The star graph (figure  \ref{fig:figure4}, left) also points to an heuristic solution of the search problem because the ideal strategy is certainly to look only at the center node and not at the others, finding the target in one try every time. But, the simple approach, ignoring the graph structure, would have a finite probability of looking at the boundary nodes. Thus, the heuristic proposed here is to start from the square root of the out-degree rule and iteratively remove nodes from consideration if such a removal improves the search time: look at a node $u$ with probability either $\sqrt{1+d_{\text{out},u}}$ or zero. Note that this is not very efficient since the setup time for a search is at least order the size of the graph so if the search only happens once, it would be just as good to just check every node. If there are many searches on a static network, then this algorithm may be advantageous, but it is more likely that the best search would combine the usual random search\cite{adamic2001} with the strategy presented here as a starting point or, for an example on the internet, to choose places to randomly jump in the popular PageRank algorithm\cite{brin1998}.

\begin{figure}[h]
\includegraphics[width=5in]{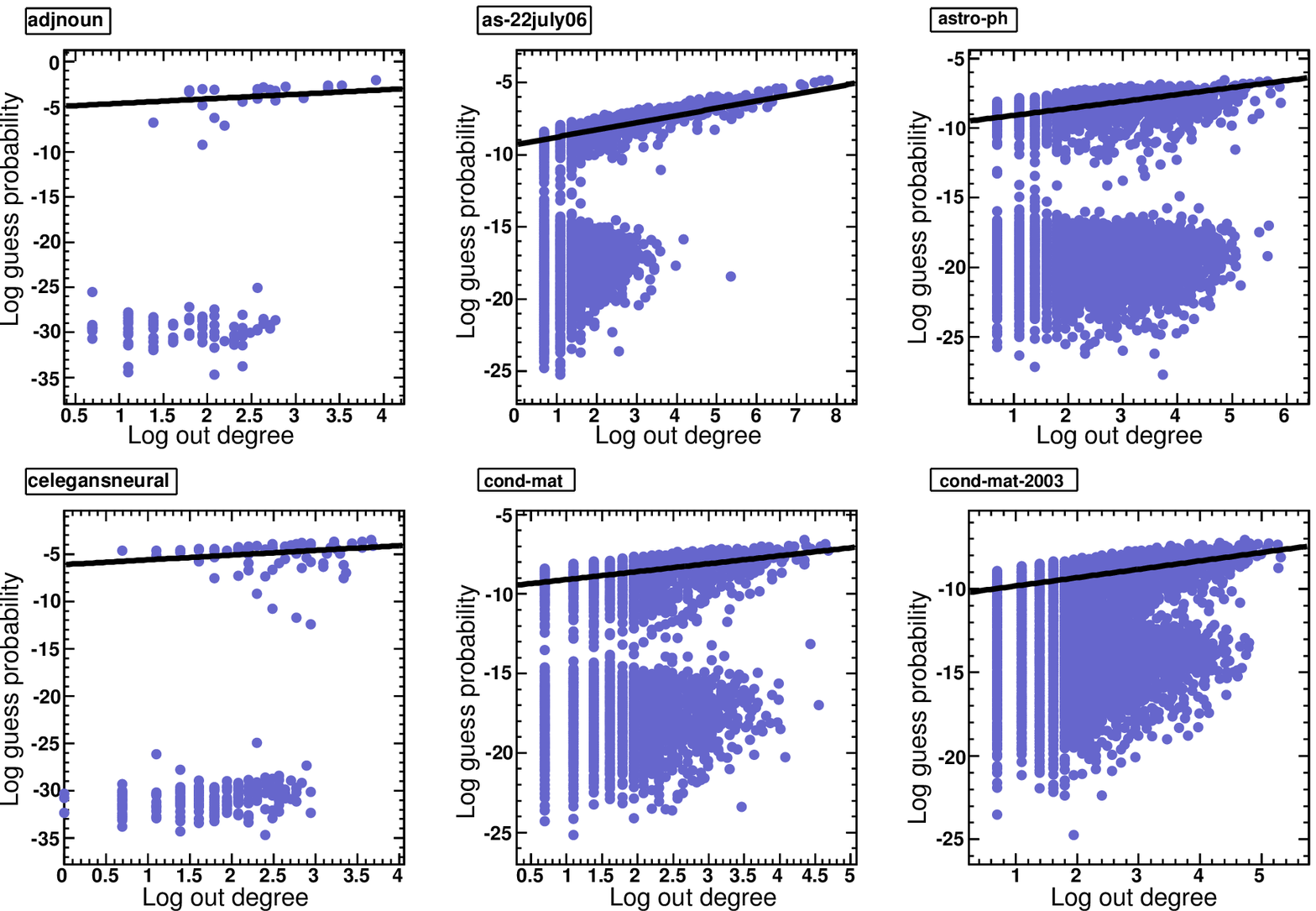}
\caption{(color online) Monte Carlo minimization of the search time for various publicly available networks. The points are the guess distribution versus the out degree plus one for each node. The line has slope 1/2. The guess probability either heads to zero (the lower cluster) or $\sqrt{1+d_{\text{out}}}$. See table \ref{tab:table1} for references.}
\label{fig:figure5}
\end{figure}

Since the proposed solution is only an heuristic, consider a direct Monte Carlo minimization (figure \ref{fig:figure5}) of the mean search time (equation \ref{eq:graph_mean_search}). The simulation starts with a random assignment of the probability to look at each node. Then, at each time step one of the nodes is selected and changed by a random amount bounded by a step size, $\delta$, and constrained to stay in the interval $\left[0,1\right]$. For a trial step, the change in mean search time is calculated from equation \ref{eq:graph_mean_search}, and steps are accepted with the usual Monte Carlo Boltzmann weight: $\exp\left(-\Delta\left<n\right>/T\right)$.  The initial step size is $\delta_{0}=0.01$, and the initial temperature is $T=0.01$ which is large compared to the usual change in probability.  After 1000 steps without finding a smaller minimum $\left<n\right>$, the temperature and step size are both lowered by a factor of two. The simulation stops after at most 10 million iterations, $T<10^{-12}$, or $\delta<10^{-12}$. Figure \ref{fig:figure5} shows the minimization for six graphs from freely available databases\cite{newman-network2006,newman-network2001,watts-network1998}. For larger graphs the convergence is poor, but the split between either the square root rule or nothing is apparent. Table \ref{tab:table1} shows the average search time for various publicly available data sets. A simple model looking at nodes with probability proportional to the out degree plus one (with no square root) is included for comparison when the Monte Carlo optimization is not feasible.

\begin{table}
\begin{ruledtabular}
\begin{tabular}{|c|c||c|c|c|}
 \hline
 Data set & Size & Probability match & Monte Carlo & Heuristic \\ 
 \hline
 www\cite{albert-network1999} & $1795408$  & $62600\pm400$ & na & $33200\pm400$ \\
 adjnoun\cite{newman-network2006} & $962$ & $17.4\pm0.3$ & $9.5\pm0.1$ &  $10.1\pm0.2$ \\
 as-22july06\cite{newman-network-nocite} & $119835$ & $2120\pm60$ & $710\pm10$ & $760\pm20$ \\
 astro-ph\cite{newman-network2001} & $258548$ & $6200\pm200$ & $1220\pm20$ & $1420\pm20$ \\
 celegansneural\cite{watts-network1998} & $2642$ & $40.1\pm0.7$ & $35.4\pm0.5$ & $41.1\pm0.7$ \\
 cond-mat\cite{newman-network2001} & $111452$ & $4700\pm100$ & $1930\pm24$ & $2120\pm30$ \\
 cond-mat-2003\cite{newman-network2001} & $270518$ & $9000\pm200$ & $2850\pm40$ & $3180\pm40$ \\
 cond-mat-2005\cite{newman-network2001} & $390963$ & $11000\pm300$ & na & $3830\pm60$ \\
 hep-th\cite{newman-network2001} & $39112$ & $2710\pm14$ & na & $1440\pm20$ \\
 netscience\cite{newman-network2006} & $6945$ & $585\pm3$ & na & $316\pm4$ \\
 polblogs\cite{Adamic05thepolitical} & $20246$ & $234\pm3$ & $195\pm1$ & $222\pm4$ \\
 power\cite{watts-network1998} & $18129$ & $1518\pm6$ & na & $1210\pm10$ \\
 % & & & & \\
 \hline
\end{tabular} 
\end{ruledtabular}
\caption{This is a table of the average search time for a randomly hidden target on the specified network. Probability match looks at nodes with probability proportional to their out degree plus one as a default comparison, Monte Carlo is a minimization as described in the text, and heuristic is the square root or nothing heuristic described in the text. Errors are all estimated by running the search $10,000$ times for the calculated guess distribution and a random target.}
\label{tab:table1}
\end{table}

\section{Conclusion}

In conclusion, a searcher that has random access to possible target positions and is trying to locate only a single target should sample a guess distribution as described here. If the searcher must be very close to a target continuously distributed in any dimension to locate it, then the guess distribution is approximately the square root of the target distribution. Otherwise, the guess distribution is more difficult to estimate, but can be numerically evaluated with sufficient care. On discrete networks an heuristic approach is to either look at a node with probability proportional to the square root of its out degree plus one or not at all. Determining a fast, local algorithm to decide whether or not to look at a node remains, but the search times from the heuristic compare favorably to those from a Monte Carlo estimation of the optimal search. 

\begin{acknowledgments}
I thank Charles F. Stevens and Ruadhan O'Flanagan for helpful comments and discussion, and Leanne Chukoskie for the experimental inspiration. This is supported in part by ONR MURI Award No.:  N00014-10-1-0072.
\end{acknowledgments}

\newpage
%\bibliography{scibib}
%Merlin.mbs v4.21 2009-07-09.
%
\end{document}